\documentclass[12pt,a4paper]{article}
\usepackage{amssymb,amsmath}
\usepackage[T1]{fontenc}
\usepackage[utf8]{inputenc}
\usepackage{authblk}
\usepackage[title,titletoc,toc]{appendix}

\title{Random matrix theory and critical phenomena in quantum spin chains}
\author[1]{J. Hutchinson\footnote{j.hutchinson@bristl.ac.uk}}
\author[1]{J. P. Keating\footnote{j.p.keating@bris.ac.uk}}
\author[1]{F. Mezzadri\footnote{f.mezzadri@bris.ac.uk}}
\affil[1]{School of Mathematics, University of Bristol, Bristol BS8 1TW, UK}

\begin{document}

\maketitle

\begin{abstract}
We compute critical properties of a general class of quantum spin chains which are quadratic in the Fermi operators and can be solved exactly under certain symmetry constraints related to the classical compact groups $U(N)$, $O(N)$ and $Sp(2N)$. In particular we calculate critical exponents $s$, $\nu$ and $z$, corresponding to the energy gap, correlation length and dynamic exponent respectively. We also compute the ground state correlators $\left\langle  \sigma^{x}_{i} \sigma^{x}_{i+n} \right\rangle_{g}$,  $\left\langle \sigma^{y}_{i} \sigma^{y}_{i+n} \right\rangle_{g}$ and  $\left\langle \prod^{n}_{i=1} \sigma^{z}_{i} \right\rangle_{g}$, all of which display quasi-long-range order with a critical exponent dependent upon system parameters. Our approach establishes universality of the exponents for the class of systems in question.
\end{abstract}

Statistical mechanical models, such as the Ising model \cite{2disingmccoywu}, enable the theoretical study of critical points and phase transitions; for example, those in magnetic systems. It is believed, though not yet rigorously proved, that such transitions have universal features, with asymptotic singularities defined by critical exponents whose values are the same within wide classes of systems. These exponents are expected to depend only on global characteristics, such as symmetry and dimensionality. This is the case for both classical and quantum phase transitions \cite{sachdev,sondhi}. It is the symmetry dependence which we seek to expose in this paper. 
 
Previously, critical exponents have been calculated rigorously for some exactly solvable models \cite{baxter}, and Renormalisation Group techniques \cite{cardy} have been developed to compute them in general classes of systems and to explain their universality. However, understanding universality in a mathematically rigorous way remains a major open problem.  We seek to address this problem here with the use of techniques from random matrix theory, which provide a natural environment for studying symmetry-related phenomena.

Random matrix theory \cite{oxrmt} has found applications in many areas of mathematical physics ranging from quantum gravity \cite{qugrav}, quantum chaos \cite{quch} and optics \cite{quopt}, to entanglement problems in quantum spin chains \cite{mezzlong, jinkore, kmlong, kmshort, kmn, klw}. In particular, for translationally invariant spin systems (e.g. the XX model), the entanglement entropy can be expressed as the determinant of a Toeplitz matrix \cite{kmlong, kmshort}. Such matrices are important in many areas of Physics, including Osanger's calculation of the diagonal spin-spin correlation of the two-dimensional classical Ising model in 1946; see, for example, the review by  Deift, Its and Krasovsky \cite{toe}. The connection with random matrices then comes about because averages over the unitary group $U(N)$ (often referred to in the Physics literature as the Circular Unitary Ensemble or CUE) may also be expressed as Toeplitz determinants.

The study of entanglement in spin chains was extended in \cite{kmlong, kmshort} to encompass Hamiltonians which possess symmetries characterising that of the Haar measure of each of the classical compact groups $U(N)$, $O^{+}(2N)$, $Sp(2N)$, $O^{+}(2N+1)$, $O^{-}(2N+1)$ and $O^{-}(2N+2)$.  This can be thought of as an example of the general classification of quantum systems developed by Altland and Zirnbauer \cite{altzin1,altzin2,zin3}. In each case it was found that the entanglement could be expressed as the determinant of matrices with combinations of Toeplitz plus Hankel structures, the asymptotics of which can be computed using proven instances of the Fisher-Hartwig conjecture and its generalisations. In all cases, it was found that changes in the analytic properties of the Fisher-Hartwig symbol of these matrices are the signature of phase transitions. The entanglement of a system and its critical properties are closely related, since when discussing the ground state (also a pure state), any correlations must be a due to entanglement.  It is this relationship that we seek to exploit now. 

We here make use of the results from \cite{kmlong, kmshort}, recapitulated in Section~\ref{sec:genqu}, extending the application of random matrix theory to compute critical properties of the same general class of quantum spin chains as was considered in \cite{kmlong, kmshort}, enabling us to expose an explicit dependence of critical behaviour on system symmetries, and establishing universality of the critical exponents for this class of systems. In particular, in Section~\ref{sec:enspec} we compute the critical exponents $s$, $\nu$ and $z$, which are related to the energy gap, correlation length and dynamic exponent respectively, and in Section 3 we compute the ground state correlators $\left\langle  \sigma^{x}_{i} \sigma^{x}_{i+n} \right\rangle_{g}$,  $\left\langle \sigma^{y}_{i} \sigma^{y}_{i+n} \right\rangle_{g}$ and  $\left\langle \prod^{n}_{i=1} \sigma^{z}_{i} \right\rangle_{g}$, which exhibit quasi-long-range order behaviour when the quantum system is gapless, decreasing as a power law of the distance $n$, with an exponent dependent upon the symmetry class.  In a companion paper to this \cite{hkm}, we then use the mapping between 1-D quantum spin chains and 2-D classical spin models to extend our results to classical systems.


\section{A class of exactly solvable one-dimensional quantum spin chains.}
\label{sec:genqu}

We consider the general class of one-dimensional quantum systems of $M$ spin-1/2 particles in an external field $h$ given by the following Hamiltonian quadratic in the Fermi operators $b_{j}$:
\begin{equation}
\mathcal{H} = \sum^{M}_{j,k} \left( A_{jk} b^{\dagger}_{j} b_{k} + \frac{\gamma}{2} B_{jk} \left( b^{\dagger}_{j} b^{\dagger}_{k}- b_{j} b_{k} \right) \right) -   2 h \sum^{M}_{j=1} b^{\dagger}_{j} b_{j},
\label{generalh}
\end{equation}
where the $b_{j}$s satisfy the usual Fermi commutation relations 
\begin{equation}
\left\{b^{\dagger}_{j}, b_{k}\right\} = \delta_{j,k}, \quad \left\{b^{\dagger}_{j},b^{\dagger}_{k}\right\} = \left\{b_{j},b_{k}\right\} = \left(b^{\dagger}_{j} \right)^{2} = \left(b_{j}\right)^{2} = 0.
\label{fermicomm}
\end{equation}
Periodic boundary conditions $b_{M+j} = b_{j}$ are assumed, and the measure of anisotropy $\gamma$ is real, with $0 \leq \gamma \leq 1$. The matrix $A_{j,k}$ must be Hermitian and $B_{j,k}$ antisymmetric, and without loss of generality we will consider them both to contain only real parameters.

It is known \cite{lieb} that such a model can be exactly diagonalised such that \eqref{generalh} becomes
\begin{equation}
\mathcal{H} =  \sum_{q} \left| \Lambda_{q} \right| \eta^{\dagger}_{q} \eta_{q} + C,
\label{hdiag}
\end{equation}
where $\left| \Lambda_{q} \right|$ is the dispersion relation determined by $A_{j,k}$ and $B_{j,k}$, $\eta_{q}$ are Fermi operators and $C$ is a constant given by
\begin{equation}
C= \frac{1}{2} \sum^{M}_{q=1} \left( A_{qq} - 2h - \left| \Lambda_{q} \right| \right).
\end{equation}

Restricting the Hamiltonian \eqref{generalh} to possess symmetries characterising the Haar measure of each of the classical compact groups $U(N)$, $O^{+}(2N)$, $Sp(2N)$, $O^{+}(2N+1)$, $O^{-}(2N+1)$ and $O^{-}(2N+2)$\footnote{for an introduction see \cite{kmlong}}, Keating and Mezzadri obtained \cite{kmlong, kmshort} 
\begin{equation}
\begin{aligned}
\Lambda_{q} & = 2^{p+1}  \left( \Gamma + \sum^{L}_{k=1} \left( a(k) \cos k q + i b(k) \sin k q \right) \right) \\
& = 2^{p+1} \left( a_{q} + i b_{q} \right),
\label{lambda}
\end{aligned}
\end{equation}
with real and imaginary parts\footnote{Note for all symmetry classes other than $U(N)$, $\gamma=0$ and thus $b_{q} = b(k)=0$, and $\Lambda_{q}$ is real.} given by 
\begin{equation}
a_{q} = \Gamma + \sum^{L}_{k=1} a(k) \cos k q, \quad \mbox{and} \quad b_{q} = \sum^{L}_{k=1} b(k) \sin k q,
\label{aborig}
\end{equation}
and 
\begin{equation} 
\Gamma  = \frac{1}{2}
\begin{cases}
a(0), \quad L = \frac{M-1}{2}, & \mbox{if M is odd}, \\
a(0) + (-1)^{l} a(\frac{M}{2}) , \quad L = \frac{M}{2}-1, & \mbox{if M is even}, \\
\end{cases}
\end{equation}
with
\begin{equation}
p =
\begin{cases}
0 & \mbox{for $U(N)$ symmetry}, \\
1 & \mbox{for all other symmetry classes}.
\end{cases}
\label{ph}
\end{equation}
Here $q$ is the wave number
\begin{equation}
q = \frac{2 \pi l}{M},
\end{equation}
with
\begin{equation}
l = 0, \ldots, M-1,
\end{equation}
for translationally invariant systems\footnote{For the other symmetry classes see \cite{kmlong}.}.

These symmetry constraints were achieved using real functions $a(j)$ and $b(j)$, even and odd functions of $\mathbb{Z}/M\mathbb{Z}$ respectively, to dictate the entries of matrices $A_{j,k}$ and $B_{j,k}$, as reported in Table~\ref{tab:ccgg} in Appendix~\ref{sec:appsymcl}.

\section{Energy spectrum and critical exponents}
\label{sec:enspec}

From \eqref{hdiag}, we see that the class of quantum systems \eqref{generalh} is gapped whenever
\begin{equation}
\left| \Lambda_{q_{c}} \right| > 0,
\end{equation}
where $q_{c}$ is the value of $q$ for which $\left|\Lambda_{q} \right|$ has an absolute minimum, and the energy gap $\Delta$ is given by 
\begin{equation}
\Delta = \left| \Lambda_{q_{c}} \right|.
\end{equation}

A signature of a continuous quantum phase transition \cite{sachdev,sondhi} for a gapped system is that the energy gap $\Delta$ vanishes as the critical point is approached according to the power law
\begin{equation}
\Delta \sim \left|g-g_{c}\right|^{s},
\label{deltaqu}
\end{equation}
where $s$ is the mass gap exponent. This behaviour holds as $g \rightarrow g_{c}$ from above or below, where $g$ is the parameter driving the phase transition (such as external field) with critical value $g_{c}$.

In addition, as the critical point is approached, we expect the divergence of the characteristic length scale $\xi$ \cite{sachdev,sondhi} to take the form 
\begin{equation}
\xi \sim \left|g-g_{c}\right|^{-\nu},
\label{qucorrspace}
\end{equation}
where $\nu$ is the \textit{correlation length critical exponent}. This length scale is often the correlation length, determining the exponential decay of correlations. 

The ratio between the two exponents in \eqref{deltaqu} and \eqref{qucorrspace} is called the \textit{dynamic critical exponent} $z$ \cite{sachdev,sondhi}, 
\begin{equation}
z = \frac{s}{\nu}.
\label{dyz}
\end{equation}
 
In addition, the dynamic exponent governs the vanishing of the dispersion relation (the energy spectrum in momentum space) $\left| \Lambda_{q} \right|$ as a function of $q$ \cite{zexp,bikasnew,chdutta};
\begin{equation}
\left| \Lambda_{q \rightarrow 0} \right| \sim q^{z},
\label{vanishdisper}
\end{equation}
where $q$ is a label in momentum space. If the energy gap $\Delta$ is also given by $\left| \Lambda_{q = 0} \right|$, then the dynamical critical behaviour at the point $q=0$ has the scaling form given by \cite{zexp,bikasnew,chdutta},
\begin{equation}
\left|\Lambda_{q \rightarrow 0} \right| \sim q^{z} \left(1+ \left(q \xi \right)^{-z} \right), \quad \mbox{as } q \rightarrow 0.
\label{scalingform}
\end{equation} 

To compute these exponents $s$, $\nu$ and $z$ for our class of quantum spin chains \eqref{generalh} we consider the following cases.

\subsection{$\gamma =0$}
\label{sec:g0}

Assuming that $\gamma = 0$, thus considering all symmetry classes, we find that $\Lambda_{q}$ is a real valued function of the external parameter $\Gamma$.

In this case, the energy gap is given by
\begin{equation}
\Delta = \left| \Lambda_{q_{c}} \right| = 2^{p+1} \left| \Gamma +  \sum^{L}_{k=1}  a(k) \cos k q_{c}  \right|,
\label{deltageniso}
\end{equation}
which vanishes (the system becomes gapless) at a critical value $\Gamma_{c}$ satisfying
\begin{equation}
\begin{aligned}
\Gamma_{c} = - \sum^{L}_{k=1}  a(k) \cos k q_{c}.
\label{critgeniso}
\end{aligned}
\end{equation}
Comparing with \eqref{deltaqu}, we find that $s=1$ for this critical point \eqref{critgeniso}, and from \eqref{dyz} we have the following relationship between the dynamic critical exponent $z$ and the correlation length exponent $\nu$
\begin{equation}
z= \frac{1}{\nu}.
\label{znu}
\end{equation}

From \eqref{vanishdisper}, we see that $z$ is the exponent governing the vanishing of $\left|\Lambda_{q} \right|$ as a function of $q$, thus for the critical points given by \eqref{critgeniso}, where the value of $q=q_{c}$ corresponds to an extremum, we will always have even $z$.

For example, any system with 
\begin{equation}
\sum^{L}_{k=1} a(k) k^{2} \cos k q_{c} \neq 0
\end{equation}
will belong to the same universality class with $z=2$ and $\nu = \frac{1}{2}$. An example is the well known nearest neighbour ($L=1$) quantum XX model (isotropic XY model).


\subsection{$\gamma \neq 0$}

For $\gamma \neq 0$ we see from Table~\ref{tab:ccgg} in the Appendix that we are restricted to systems with $U(N)$ symmetry only. In this situation the energy gap is given by
\begin{equation}
\Delta = \left| \Lambda_{q_{c}} \right| = 2^{p+1} \sqrt{ \left( \Gamma +  \sum^{L}_{k=1}   a(k) \cos k q_{c} \right)^{2} + \left( \sum^{L}_{k=1} b(k) \sin k q_{c} \right)^{2}}.
\label{deltageniso}
\end{equation}
There are now several possibilities.

First, assuming $q_{c}=0, \pi$ we recover the results from Section~\ref{sec:g0}. Alternatively if $\gamma =1$ and $a(k) = b(k)$ for all $k$, \eqref{deltageniso} becomes
\begin{equation}
\Delta  = \left| \Lambda_{q_{c}} \right| = 2^{p+1} \sqrt{ \Gamma^{2} + \sum^{L}_{k=1} \left(a^{2}(k) + 2 a(k) \left(\Gamma \cos k q_{c} + \sum^{L-k}_{l=1} a (k+l) \cos l q_{c}  \right) \right)}. 
\label{gapgn0}
\end{equation}
In this case, any system with 
\begin{equation}
\sum^{L}_{k=1} \left(a^{2}(k) + 2 a(k) \left(\Gamma \cos k q_{c} + \sum^{L-k}_{l=1} a (k+l) \cos l q_{c}  \right) \right) = 2 \Gamma c + c^{2},
\end{equation}
where $c$ is a constant dependent upon the interaction coefficients $a(k)$, will belong to the same universality class, and all such systems have $z=\nu = 1$. An example is the well known nearest neighbour ($L=1$) quantum Ising model (anisotropic XY model) where $q_{c} = \pi$ and $c=a(1)$ .


\section{Correlator}
\label{sec:newcorr}

In this section we compute the following correlators for the general class of quantum spin chains \eqref{generalh} restricted to $U(N)$ symmetry only and with $\gamma =0$, in the limit $n \rightarrow \infty$:
\begin{equation}
\begin{aligned}
& \left\langle \prod^{i+n}_{l=i} \sigma^{z}_{l} \right\rangle_{g} = \left( -1 \right)^{n} \det{\mathcal{M}_{n}}, \\
& \left\langle \sigma^{x}_{i} \sigma^{x}_{i+n} \right\rangle_{g} = \left(-1 \right)^{n} \det \mathcal{M}^{x}_{n},  \\
& \left\langle \sigma^{y}_{i} \sigma^{y}_{i+n} \right\rangle_{g}= \left( -1 \right)^{n} \det \mathcal{M}^{y}_{n},
\end{aligned}
\end{equation}
where $\left\langle . \right\rangle_{g}$ is the expectation value with respect to the ground state of the quantum system. 

As summarised in Table~\ref{tab:ccgg} in the Appendix, it was found in \cite{kmlong, kmshort} that for all symmetries considered, the matrices $\mathcal{M}_{n}$ are combinations of Toeplitz plus Hankel matrices
\begin{equation}
\mathcal{M}_{n} = T_{n} \pm H_{n}, 
\end{equation}
where $T_{n}$ represents a Toeplitz matrix and $H_{n}$ represents a Hankel matrix:
\begin{equation}
T_{n} \left[ g \right] = \left\{ g_{j-k} \right\}_{j,k = 0,\ldots n-1}, \qquad H_{n} \left[ g \right] = \left\{ g_{j+k+c} \right\}_{j,k = 0,\ldots n-1},
\label{toehank}
\end{equation}
where $c$ is a constant given by Table~\ref{tab:ccgg}, and $g_{l}$ are the Fourier coefficients of the symbol $g^{\mathcal{M}} \left( \theta \right)$ given by 
\begin{equation}
g^{\mathcal{M}}\left( \theta \right) = \frac{\Lambda \left( \theta \right)}{\left| \Lambda \left(\theta \right) \right|},
\label{gtsymb} 
\end{equation}
with $\Lambda \left( \theta \right)$ obtained by taking the limit $M \rightarrow \infty$ in \eqref{lambda}.

The matrices $\mathcal{M}^{x}_{n}$ and $\mathcal{M}^{y}_{n}$ can be written as the combination of Toeplitz plus Hankel matrices
\begin{equation}
\begin{aligned}
\mathcal{M}^{x}_{n} & = T^{x}_{n} \pm H^{x}_{n},\\ 
\mathcal{M}^{y}_{n} & = T^{y}_{n} \pm H^{y}_{n},\\ 
\end{aligned}
\end{equation}
where $T^{x}_{n}$ and $T^{y}_{n}$ represent Toeplitz matrices with the same structure as in \eqref{toehank}, but now the $g_{l}$s are Fourier coefficients of the symbols $g^{T^x} \left( \theta \right) = e^{i \theta} g^{\mathcal{M}} \left( \theta \right)$ and $g^{T^y} \left( \theta \right) = e^{-i \theta} g^{\mathcal{M}} \left( \theta \right)$ respectively.  Similarly $H^{x}_{n}$ and $H^{y}_{n}$ represent Hankel matrices with entries given by Fourier coefficients of the symbols $g^{H^x} \left( \theta \right) = g^{H^y} \left( \theta \right) = e^{-i \theta} g^{\mathcal{M}} \left( \theta \right)$.

When $\gamma =0$, $\mathcal{M}_{n}$ is a symmetric matrix with symbol $g^{\mathcal{M}}\left( \theta \right)$ given by \eqref{gtsymb}. In this case when the system is gapped and away from the critical point\footnote{That is the external field $\left|\Gamma \right| > \left| \Gamma_{c} \right|$, thus all the spins are aligned in the direction of the field.}, $g^{\mathcal{M}}\left( \theta \right)$ is a constant taking values $\pm 1$, and $\mathcal{M}_{n} = \pm I$.  Therefore we find that
\begin{equation}
\left\langle \prod^{i+n}_{l=i} \sigma^{z}_{l} \right\rangle_{g} = \pm 1, \quad \mbox{and} \quad \left\langle \sigma^{x}_{i} \sigma^{x}_{i+n} \right\rangle_{g} = \left\langle \sigma^{y}_{i} \sigma^{y}_{i+n} \right\rangle_{g} = 0.
\label{asaway}
\end{equation}

When the system is gapless, $g^{\mathcal{M}} \left( \theta \right)$ is a piece-wise continuous even function taking values $1$ and $-1$ and has discontinuities at all points $\theta_{r}$ satisfying 
\begin{equation}
\Lambda \left( \theta_{r} \right) =0,
\label{rsolns}
\end{equation}
with the additional condition that the the first non-zero derivative of $\Lambda \left( \theta \right)$ at $\theta_{r}$ is odd.

Symbols of this type can be written in Fisher-Hartwig form:
\begin{equation}
g^{\mathcal{M}} \left( \theta \right) = \phi \left( \theta \right) \prod^{L}_{r=1} u_{\alpha_{r}} \left( \theta - \theta_{r} \right) t_{\beta_{r}} \left( \theta- \theta_{r} \right),
\label{fhsymbform} 
\end{equation}
where $\phi$ is smooth, has winding number zero and
\begin{equation}
\begin{aligned}
t_{\beta} \left( \theta \right) & = e^{-i \beta \left( \pi - \theta \right)}, \quad 0 \leq \theta < 2 \pi, \quad \beta \notin \mathbb{Z}, \\
u_{\alpha} \left( \theta \right) & = \left(2-2\cos \theta \right)^{\alpha}, \quad \operatorname{Re} \alpha > - \frac{1}{2},
\end{aligned}
\end{equation}
and $L$ is the number of zeros/discontinuities in the interval $[0,2 \pi)$. Note here that the term $\left( 2 - 2 \cos \theta \right)^{\alpha}$ has a zero if $\operatorname{Re} \alpha > 0$, and a pole if $\operatorname{Re} \alpha < 0$, and an oscillating discontinuity if $\operatorname{Re} \alpha = 0$ but $\operatorname{Im} \alpha_{r} \neq 0$. The term $e^{i \beta \left( \theta - \pi \right)}$ is a function with a jump discontinuity with limit $e^{-i\beta \pi}$ ($e^{i \beta \pi}$) as $\theta \rightarrow + 0$ ($\theta \rightarrow 2\pi-0$).

In our case $g^{\mathcal{M}} \left(\theta \right)$ is even and has only discontinuities, hence \eqref{fhsymbform} can be simplified further as
\begin{equation}
g^{\mathcal{M}} \left(\theta \right) = \phi \left( \theta \right) \prod^{R}_{r=1} t_{\beta_{r}} \left( \theta - \theta_{r} \right) t_{-\beta_{r}} \left( \theta + \theta_{r} \right),
\label{symbform}
\end{equation}
where now $R$ is the number of discontinuities in the interval $\left[0,\pi \right)$ (thus $L=2R$ when comparing to \eqref{fhsymbform}). 

We now make use of the generalised Fisher-Hartwig Theorem, conjectured by Basor and Tracy \cite{ebct} and proved by Deift, Its and Krasovsky \cite{dikfh} for $\beta_{r} \in  \mathbb{C}$:
\begin{equation}
\det T_{n} [g] \sim \sum_{\mbox{\tiny Rep.}} e^{c_{0} n} n^{- \sum^{L}_{r=1} \beta^{2}_{r}}E, \quad n \rightarrow \infty,
\label{fhg}
\end{equation}
where $c_{0}$ is the zeroth Fourier coefficient of $\ln \phi \left( \theta \right)$, the sum is over the different representations of \eqref{symbform} corresponding to the minimum exponent $\sum_{r} \beta^{2}_{r} $, and $E$ is a constant given by \cite{dikfh,kmlong}
\begin{equation}
\begin{aligned}
E & = e^{\sum^{\infty}_{k=1} k c_{k} c_{-k}} \prod^{L}_{r=1} \left( \phi_{+} \left( e^{i \theta_{r}} \right) \right)^{- \beta_{r}} \left( \phi_{-} \left( e^{-i \theta_{r}} \right) \right)^{\beta_{r}} \\
& \quad \times \prod_{0 \leq r \neq s \leq L} \left(1-e^{i \left(\theta_{s} - \theta_{r} \right)} \right)^{\beta_{r}  \beta_{s}} \prod^{L}_{r=1} G\left(1- \beta_{j} \right) G\left(1 + \beta_{j} \right),
\end{aligned} 
\end{equation}
where $G(x)$ is the Barnes' G-function and
\begin{equation}
\ln \phi_{+} \left(t \right) = \sum^{\infty}_{j=1} c_{j} t^{j}, \quad \ln \phi_{-} \left(t \right) = \sum^{\infty}_{j=1} c_{-j} t^{-j}.
\end{equation} 

In our case we have $2^{R}$ representations for \eqref{symbform}, all with the form\footnote{This representation assumes that \eqref{gtsymb} is $+1$ in the region surrounding $\theta=0$.} 
\begin{equation}
\phi \left( \theta; \pm \theta_{1}, \ldots, \pm \theta_{R} \right) = \left(-1 \right)^{R} e^{i \sum^{R}_{r=1} \pm \theta_{r}} \quad \mbox{with} \quad \beta_{r} = \pm \frac{1}{2}, 
\label{symparam}
\end{equation}
where the sign in front of $\beta_{r}$ matches the sign in front of $\theta_{r}$ and the $2^{R}$ representations correspond to the $2^{R}$ possible sign combinations for each $\theta_{r}$. 

Until now \eqref{fhg} has been proven for determinants of Toeplitz matrices only, thus the following results only apply to our class of quantum systems \eqref{generalh} corresponding to $U(N)$ symmetry. As $n \rightarrow \infty$ we obtain
\begin{equation}
\begin{aligned}
&  \left\langle \prod^{i+n}_{l=i} \sigma^{z}_{l} \right\rangle_{g} = n^{-\frac{R}{2}} \left(-1 \right)^{n\left(R+1\right)} F \sum_{\mbox{\tiny Rep.}} \mathcal{Q}, \\
& \left\langle \sigma^{x}_{i} \sigma^{x}_{i+n} \right\rangle_{g} =  n^{-\frac{R}{2}} \left(-1 \right)^{R n} F^{x} \sum_{\mbox{\tiny Rep.}}  \mathcal{Q}, \\
& \left\langle \sigma^{y}_{i} \sigma^{y}_{i+n} \right\rangle_{g} = n^{-\frac{R}{2}} \left(-1 \right)^{R n} F^{y} \sum_{\mbox{\tiny Rep.}} \mathcal{Q}, 
\label{asnear} 
\end{aligned}
\end{equation}
when the systems is gapless, with constants
\begin{equation}
\begin{aligned}
F & = \left(G\left(\frac{1}{2}\right) G\left(\frac{3}{2} \right)\right)^{2R} \prod^{R}_{r=1} \left| 1-e^{i 2 \theta_{r}} \right|^{-\frac{1}{2}}, \\
F^{x} & = e^{-\sum^{\infty}_{k=1} \frac{1}{k} } F \quad \mbox{and} \quad F^{x}= e^{\sum^{\infty}_{k=1} \frac{1}{k} } F.
\end{aligned}
\end{equation}
The term $\mathcal{Q}$ contains the factors which have a dependence on the representation of $\phi \left( \theta \right)$, given by
\begin{equation}
\mathcal{Q} \left( \pm \theta_{1}, \ldots \pm \theta_{R} \right)  = e^{i n \sum^{R}_{r=1} \pm \theta_{r}}  \prod_{1 \leq r < s\leq R} \left|\frac{1-e^{i \left( \theta_{r} - \theta_{s} \right)}}{1-e^{i \left( \theta_{r} + \theta_{s} \right)}} \right|^{4 \beta_{s} \beta_{r}},
\end{equation}
since in our case we have
\begin{equation}
\begin{aligned}
\prod_{0 \leq r \neq s \leq L} & \left(1-e^{i \left(\theta_{s} - \theta_{r} \right)} \right)^{\beta_{r}\beta_{s}} \\
& \qquad = \prod^{R}_{r=1} \left| 1-e^{i 2 \theta_{r}} \right|^{-\frac{1}{2}}  \prod_{1 \leq r < s\leq R} \left|\frac{1-e^{i \left( \theta_{r} - \theta_{s} \right)}}{1-e^{i \left( \theta_{r} + \theta_{s} \right)}} \right|^{4 \beta_{s} \beta_{r}}.
\end{aligned}
\end{equation}

Thus we find that away from the critical point $\left| \Gamma \right| > \Gamma_{c}$, the asymptotics \eqref{asaway} are given by a constant, whereas below the critical external field $\left|\Gamma \right| < \Gamma_{c}$ (when the system is gapless), the correlator decreases like a power law \eqref{asnear} as $n \rightarrow \infty$. Of particular interest is the fact that the exponent describing this power law depends upon the number of discontinuities of the symbol \eqref{gtsymb}, which for a specific system with fixed interaction coefficients will depend on the external field $\Gamma$. This is reminiscent of the behaviour of a \textit{Kosterlitz-Thouless transition}. 

We see that of central importance to this work is the analytic behaviour of the symbol of the matrix determinant obtained for the correlation functions and that it is the symmetries of the systems which shape the behaviour of this symbol. It is by using this symbol and its analytic properties that we are able to show how symmetries affect the critical properties.


\section{Acknowledgements}
We are grateful to Professor Shmuel Fishman for helpful discussions. JH is pleased to thank Nick Jones for several insightful remarks, to the EPSRC for support during her PhD, and to the Leverhulme Trust for further support. FM was partially supported by EPSRC research grants EP/G019843/1 and EP/L010305/1.

\newpage
\begin{appendices}
\section{Symmetry classes}
\label{sec:appsymcl}
\begin{table}[htbp]
\centering
\begin{tabular}{c c c}
\hline \hline
Classical compact & Structure of matrices & Matrix entries \\
 group & $\bar{A}_{j,k} \quad (\bar{B}_{j,k})$ & $\left( \mathcal{M}_{n} \right)_{j,k} $ \\
\hline \hline
$U(N)$ & $a(j-k) \quad (b(j-k))$ & $g_{j-k}, \quad j,k \geq 0$ \\
$O^{+}(2N)$ & $a(j-k)+a(j+k)$ & $g_{0} \quad$ if $j=k=0$\\
 & & $\sqrt{2} g_{l}$  if \\
& &  either $j=0, k=l$ \\
& & $\quad$ or $j=l, k=0$ \\
 & & $g_{j-k} + g_{j+k}, \quad j,k >0$ \\
$Sp(2N)$ & $a(j-k) -a(j+k+2)$ & $g_{j-k}-g_{j+k+2}, \quad j, k \geq 0$ \\
$O^{\pm}(2N+1)$ & $a(j-k) \mp a(j+k+1)$ & $g_{j-k} \mp g_{j+k+1}, \quad j,k \geq 0$\\ 
$O^{-}(2N+2)$ & $a(j-k) -a(j+k+2)$ & $g_{j-k} - g_{j+k+2}, \quad j,k \geq 0 $ \\
\hline
\end{tabular}
\caption{The structure of functions $a(j)$ and $b(j)$ dictating the entries of matrices $\mathbf{\bar{A}} = \mathbf{A}- 2 h \mathbf{I}$ and $\mathbf{\bar{B}} = \gamma \mathbf{B}$, which reflect the respective symmetry groups. The $g_{l}$s are the Fourier coefficients of the symbol $g^{\mathcal{M}} \left( \theta \right)$ of $\mathcal{M}_{M}$. Note that for all symmetry classes other than $U(N)$, $\gamma=0$ and thus $\bar{\mathbf{B}}=0$.}
\label{tab:ccgg}
\end{table}

\end{appendices}

\end{document}